\documentclass[fleqn,12pt,twoside]{article}
\usepackage{espcrc1}

\usepackage{graphicx}
\usepackage[figuresright]{rotating}
\def\msun{${\rm M}_{\odot}$}

\newcommand{\AmS}{{\protect\the\textfont2
  A\kern-.1667em\lower.5ex\hbox{M}\kern-.125emS}}

\title{
  How to reconcile the C and s-process abundances in the metal-poor star V453\,Oph ?
}

\author{Pieter Deroo\address[IvS]{Instituut voor Sterrenkunde,
        K.U.Leuven, \\ 
        Celestijnenlaan 200B, B-3001 Leuven, Belgium},
	St\'ephane Goriely\address[IAA]{Institut d'Astronomie et d'Astrophysique,
	  Universit\'e Libre de Bruxelles, \\
	  CP 226, 1050 Brussels, Belgium},
	Lionel Siess\addressmark[IAA],
        Maarten Reyniers\addressmark[IvS]
	and
	Hans Van Winckel\addressmark[IvS],
	}
       
\begin{document}

\maketitle

\begin{abstract}
 In this poster we present a detailed chemical analysis of the RV\,Tauri star of low
intrinsic metallicity, V453 Oph ([Fe/H]=$-2.2$), which shows a mild but clear s-process
surface overabundance. This result is strengthened by a relative analysis to the RV Tauri star DS Aqr 
of similar metal deficiency ([Fe/H]=$-1.6$), but without such s-process
enrichment. A remarkable result is that the s-process enrichment of V453 Oph is
\emph{not} accompanied by a measurable C enhancement. To explain the surface abundances observed
in V453 Oph, two nucleosynthesis models are considered. The first model assumes the
partial mixing of protons during the radiative interpulse phase while the second
scenario is based on the possible production of s-process elements within the
flash-driven convective pulse only. Both models can give a satisfactory
explanation for the s-process enrichment of V453\,Oph, but not the simultaneous low
C-abundance observed.
\end{abstract}

\section{Introduction}

 RV\,Tauri stars are rare supergiants which occupy the high luminosity end of the pop.
II Cepheid instability strip. They are identified on the basis of their characteristic
light curves, which show alternating deep and shallow minima with periods between 30 and
150 days. Their post-AGB nature is established on the basis of their far-IR excess due
to circumstellar dust, which is thought to be a relic of the dusty mass loss at the end of the AGB.
The genuine high luminosity nature of this RV\,Tauri class was established by the
detection of some members in Globular Clusters and the LMC. 

 Post-AGB objects are very useful in constraining the AGB chemical evolutionary models,
because the absence of strong molecular veiling allows detection of atomic transitions
from CNO up to the heavy s-process elements beyond the Ba-peak.

\section{Observations and reduction}
 High resolution spectra of V453\,Oph and DS\,Aqr were acquired with the spectrograph
FEROS, mounted on the 1.5\,m telescope at La Silla. These spectra in combination with
the latest KURUCZ model atmospheres were employed to derive accurate abundances for both
objects. The parameters for the model atmospheres were determined using spectroscopic
criteria only (using the Fe lines). The resulting model parameters are shown in Table
\ref{modelparameters}. The huge number of Fe lines (136 and 316 for V453\,Oph and DS\,Aqr respectively) ensures an accurate determination of these values.
\begin{table}
\caption{The model parameters of V453\,Oph and DS\,Aqr as determined using the Fe lines. The derived CNO abundances for both objects are also given.}\label{modelparameters}
\newcommand{\m}{\hphantom{$-$}}
\newcommand{\cc}[1]{\multicolumn{1}{c}{#1}}
\renewcommand{\tabcolsep}{0.9pc} 
\begin{tabular}{l lcll  lll}
\hline
\cc{starname}&\cc{T$_{\rm{eff}}$}&\cc{$\log(g)$}&\cc{$\zeta_{\rm{t}}$}&
\cc{[Fe/H]}&
\cc{[C/Fe]}&\cc{[N/Fe]}&\cc{[O/Fe]}\\
\hline
V453\,Oph & 6250 & 1.5 & 3.0 & $-2.23\pm0.12$ & \cc{$-0.26$} & \cc{0.74} & \cc{0.99} \\
DS\,Aqr & 5750 & 0.5 & 3.5 & $-1.62\pm0.12$  & \cc{$-0.15$} & \cc{0.32} & \cc{0.64} \\
\end{tabular}
\end{table}

\section{Abundance analysis}
For both objects we performed a very detailed chemical abundance analysis. All lines in the
spectra showing a clear and symmetric profile were measured and subsequently identified
on the basis of wavelength. For each of these lines, the most recent atomic data was
researched in the literature. To eliminate possible biasing in the atomic data, a strict relative analysis
between both objects was performed using exactly the same lines for both objects. Because of the very similar 
atmospheric parameters, these relative abundances are less sensitive to possible systematic
errors in general.

\begin{figure}[htp]
\centering
\includegraphics[height=8cm,angle=-90]{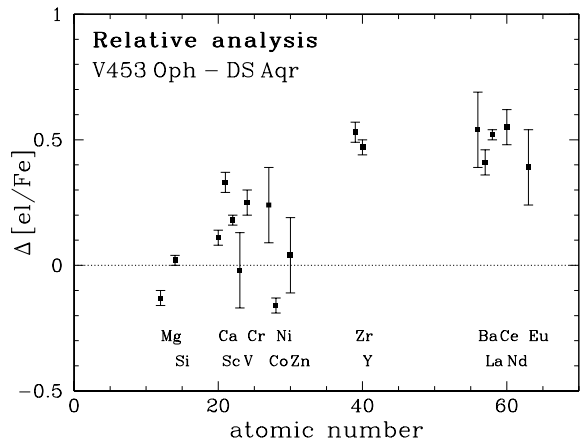}
\caption{The abundance results as determined from the relative abundance analysis.}
\label{relative_analysis}
\end{figure}

\begin{figure}
\includegraphics[height=7.85cm,angle=-90]{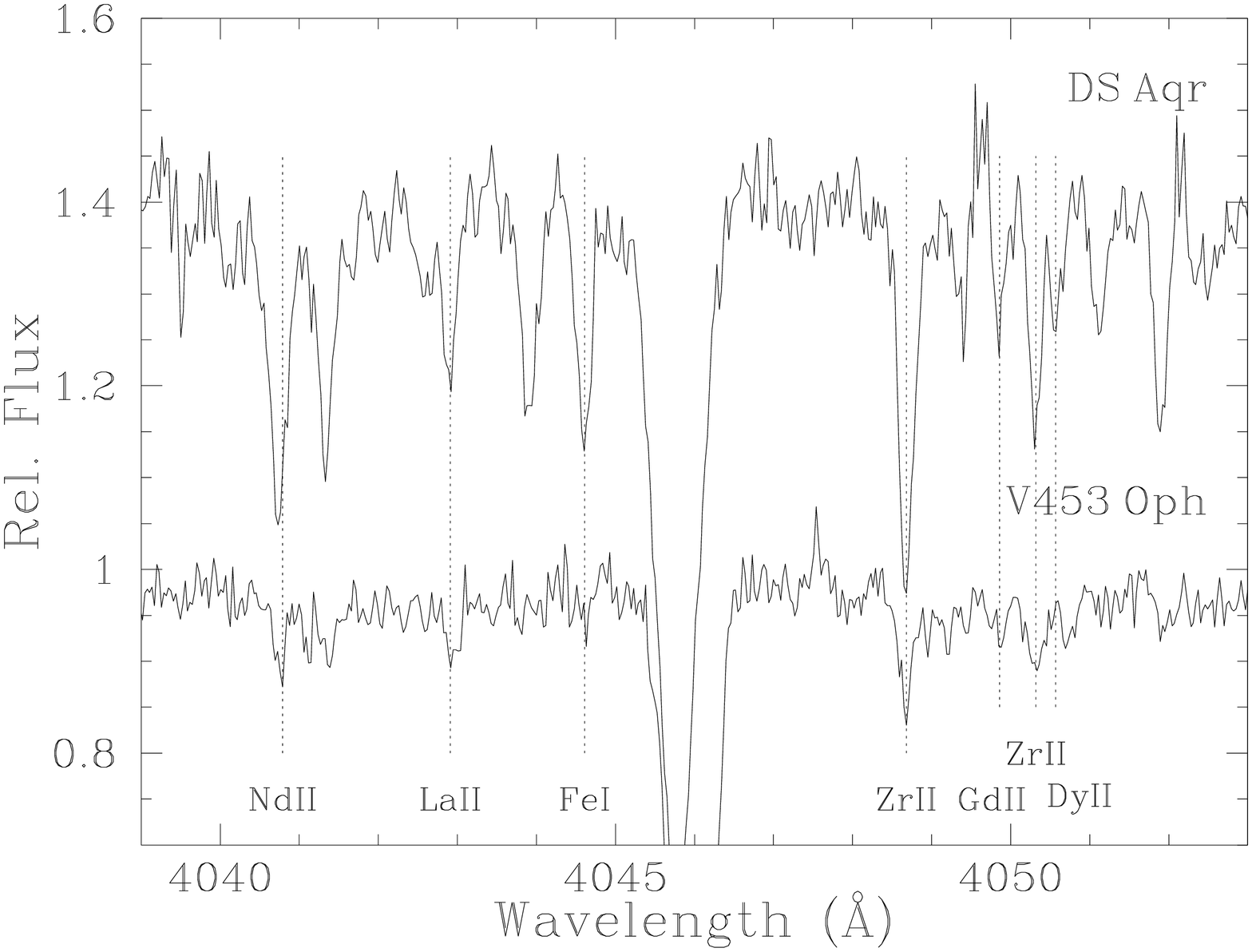}
\includegraphics[height=7.85cm,angle=-90]{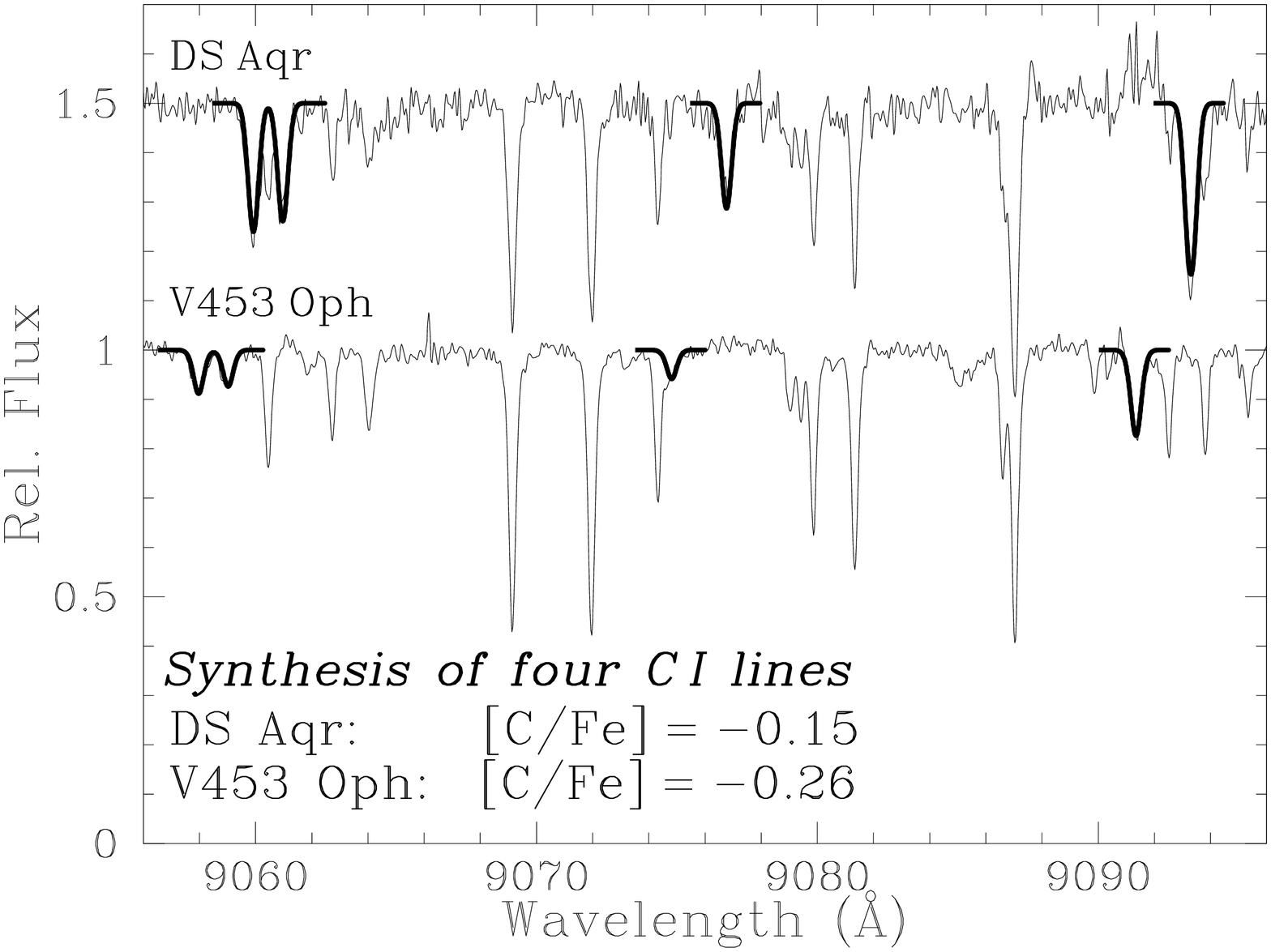}
\caption{\emph{Left panel:} Velocity corrected spectra of both objects showing some of
the employed s-process lines. \emph{Right panel:} Non-velocity corrected spectra with the
spectrum synthesis of four C\,I lines shown using a thicker line.}\label{sample_spectra}
\end{figure}

 The result of the absolute abundance analysis (i.e. for each object independently)
reveals that V453\,Oph is mildly enriched in s-process elements while DS\,Aqr is not.
Therefore, the relative abundance analysis of V453\,Oph $-$ DS\,Aqr, depicted in
Fig.\,\ref{relative_analysis}, gives extremely accurate s-process enrichment values.
Some of the s-process lines used in the analysis are shown in the left panel of
Fig.\,\ref{sample_spectra}. Although V453\,Oph is s-process enriched, the C abundance
 as derived from a spectrum synthesis (shown in the right panel of
Fig.\,\ref{sample_spectra}), is very similar to DS\,Aqr (see Table \ref{modelparameters}). 

\section{AGB nucleosynthesis}
 In general, two different scenarios are invoked to explain the s-process enrichment: in
the less massive objects, the $^{13}$C($\alpha$,n)$^{16}$O reaction, which takes place
under radiative conditions, is nowadays believed to be the major neutron source
\cite{Busso_1999}. For hot AGB stars, however, another s-process scenario
is also invoked. It concerns the production of s-elements under convective conditions by
the $^{22}$Ne($\alpha$,n)$^{25}$Mg neutron source. Both scenarios were explored in order
to explain the s-enrichment of V453\,Oph without a C enhancement. In the following, these
scenarios are referred to as radiative and convective s-process models. 

\begin{figure}
\includegraphics[width=7.85cm]{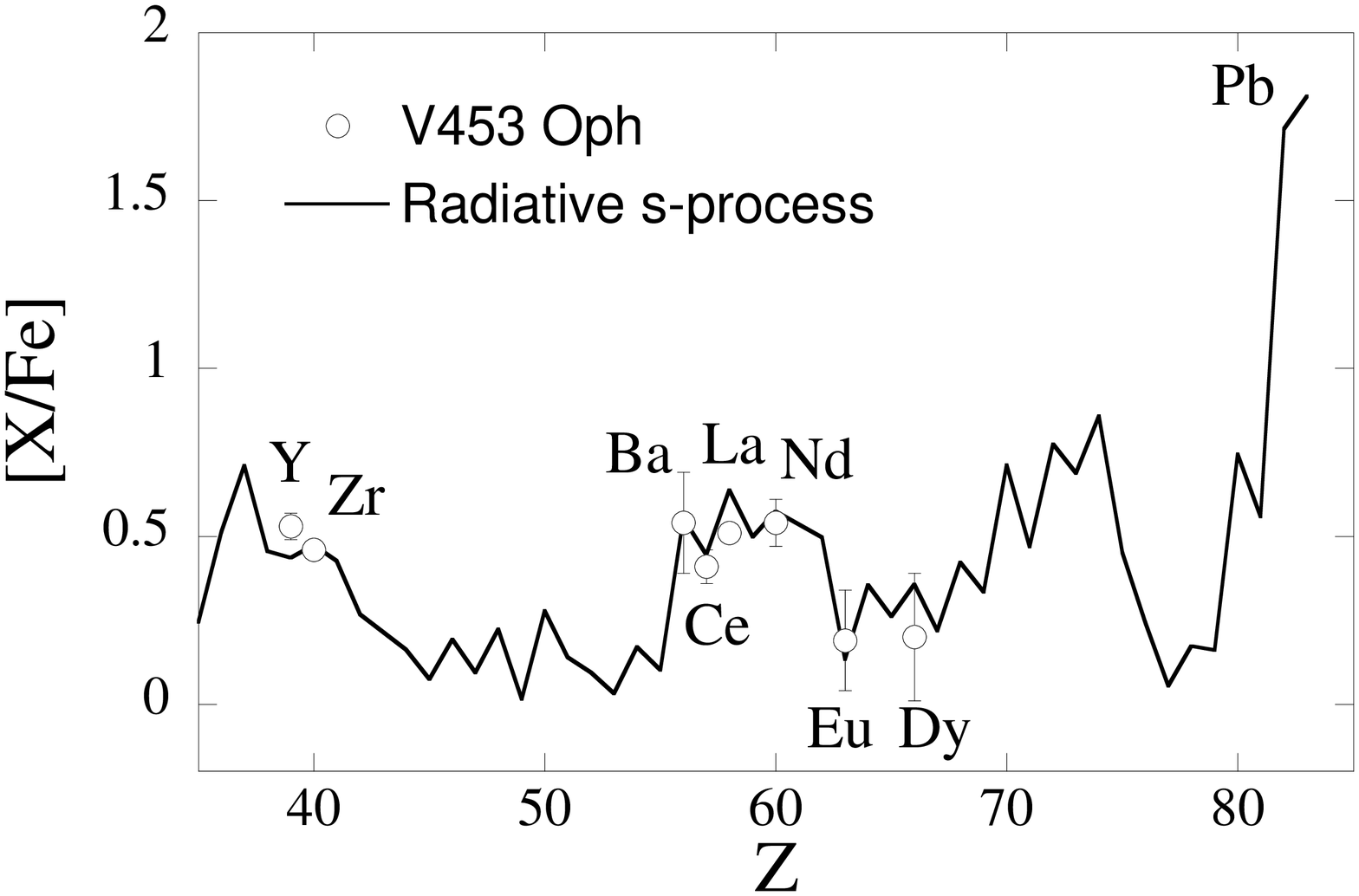}
\includegraphics[width=7.85cm]{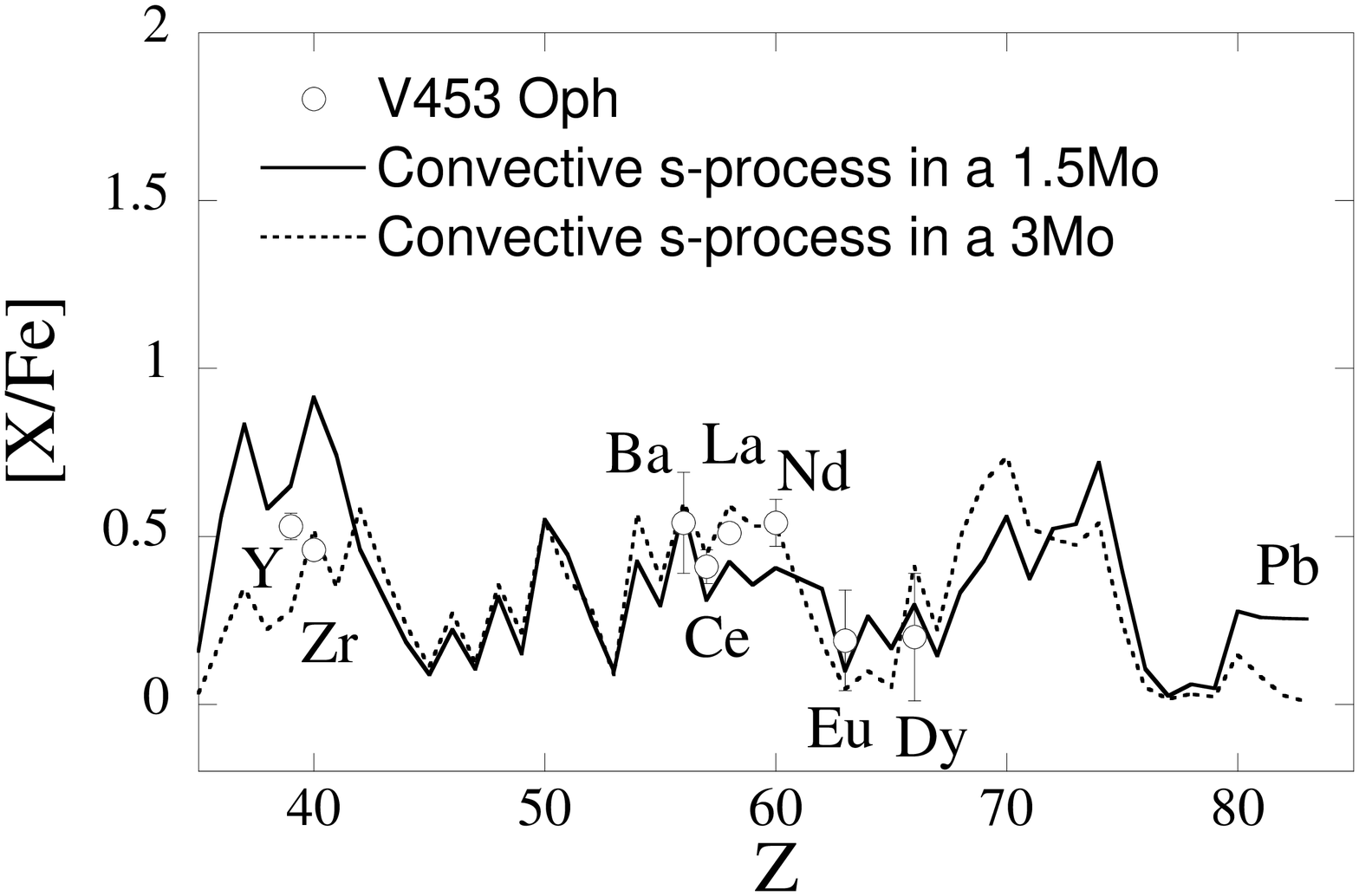}
\caption{Comparison of the observed relative surface abundances with the
radiative (\emph{left panel}) and convective (\emph{right panel}) s-process
models.}
\label{agb_enrichment}
\end{figure}

 The most favoured s-process model is associated with the partial mixing of protons
(PMP) into the radiative C-rich layers at the time of the third dredge-up (3DUP).
S-process calculations were performed using this PMP model for an 1.5\msun\ Z=0.0001 AGB
model star. The model has been evolved selfconsistently up to the 49th interpulse-pulse
sequence. However, no 3DUP is found to take place, so that only a mild s-process
develops in the convective pulse through the  $^{22}$Ne($\alpha$,n)$^{25}$Mg neutron
source. At the end of the 49th pulse, protons are artificially injected into the C-rich
layers to simulate the PMP scenario (as described in \cite{Goriely00}). An efficient
s-process develops in the radiative zone and after the 50th pulse, a 3DUP is assumed.
The left panel of Fig. \ref{agb_enrichment} compares the observed abundances with the
radiative s-process model predictions. Clearly, the abundance distribution agrees well
with the observations. However, the overabundance obtained for C ([C/Fe]$=1.5$) is
incompatible with the observed value ([C/Fe]$=-0.26$).

 In the convective model, the s-process originates exclusively from the neutron
irradiation due to the $^{22}$Ne($\alpha$,n)$^{25}$Mg source, requiring temperatures of
about 3.5 10$^8$K. In the 1.5\msun\ Z=0.0001 model star, the maximum temperature at the
bottom of the thermal pulse amounts to about 3.3 10$^8$K. 
To allow an s-process to develop in the pulse, the $^{22}$Ne($\alpha$,n)$^{25}$Mg
reaction rate has been increased by a factor of 10 with respect to the NACRE recommended
rate. This value is well within the uncertanties still affecting this rate
\cite{Koehler_2002}. With this increased rate, the full sequence of the 50
pulse-interpulse in the 1.5\msun\ Z=0.0001 model star have been calculated ending with a
parametrized dredge-up. The resulting elemental abundance distribution is shown in the
right panel of Fig.
\ref{agb_enrichment} to reproduce the observations satisfactorily. However, the C
prediction ([C/Fe]$=1.2$) is again not compatible with the observations.

 The only model that does provide a compatible C abundance with observations is a
3\msun\ Z=0.0001 model star. In this case, temperatures as high as 3.7 10$^8$K are
encountered and a significant s-process takes place. The resulting surface enrichment
obtained assuming a dredge-up at the end of the 18th computed pulse is given in Fig.
\ref{agb_enrichment} and is again in good agreement with the observations. Even the
predicted C abundance ([C/Fe]$=0.1$) is in agreement with the observations. Therefore
this seems the only model able to reproduce simultaneously the s-process enrichment and
the low C abundance. However, a 3\msun\ star with the very low metallicity of V453\,Oph
is very unlikely from evolutionary viewpoint.
\section{Conclusion}
 This paper reports upon the abundance analysis of a very particular s-process enriched
object: V453\,Oph. The abundances derived for this object are of particular
nucleosynthesis interest, since for the first time, the s-process enrichment is not
accompanied by a simultaneous C-enrichment. This observation provides us with clear
proof that our knowledge about the AGB nucleosynthesis is still unsatisfactory because no current
model can fully reproduce the specific history of this object.

\end{document}